\newcommand{\CLASSINPUTbottomtextmargin}{.67in}
\newcommand{\CLASSINPUTinnersidemargin}{.67in}
\documentclass[journal,draftcls,onecolumn,12pt,twoside]{IEEEtranTCOM}
\IEEEoverridecommandlockouts
\usepackage[utf8]{inputenc} 
\usepackage[table]{xcolor}
\usepackage{graphicx}
\graphicspath{{figs/}}

\usepackage{epstopdf}
\usepackage{pgfplots}
\pgfplotsset{compat=1.14}
\usepackage{tikz}
\usetikzlibrary{arrows, shapes}
\usepackage{tabularx}
\usepackage{dcolumn}
\usepackage{multirow}
\usepackage{float}
\usepackage{amsthm}
\usepackage[cmex10]{amsmath}
\usepackage{amssymb}
\usepackage{braket}
\usepackage{array}
\usepackage{longtable}
\usepackage{arydshln}
\usepackage{bm}
\usepackage{bbm}
\usepackage{bbold}

\usepackage{mathtools}

\newcommand{\Fig}[1]{Fig.~\ref{#1}}
\usepackage[T1]{fontenc}
\usepackage{url}
\usepackage{cite}
\usepackage[cmex10]{amsmath}

\makeatletter
\newcommand{\distas}[1]{\mathbin{\overset{#1}{\kern\z@\sim}}}%
\newsavebox{\mybox}\newsavebox{\mysim}
\newcommand{\distras}[1]{%
 \savebox{\mybox}{\hbox{\kern3pt$\scriptstyle#1$\kern3pt}}%
 \savebox{\mysim}{\hbox{$\sim$}}%
 \mathbin{\overset{#1}{\kern\z@\resizebox{\wd\mybox}{\ht\mysim}{$\sim$}}}%
}

\newcommand\ind[3]{ {#1}^{(#2)}_{#3} }
\newcommand\beg[2]{ {#1}_{[#2]}}

\newcommand{\pM}{p_{m}}
\newcommand{\pF}{p_{f}}
\newcommand{\bu}{\mathbf{u}}
\newcommand{\bv}{\mathbf{v}}
\newcommand{\bX}{\mathbf{X}}

\newtheorem{theorem}{Theorem}[section]
\newtheorem{lemma}[theorem]{Lemma}

\newtheorem{define}{Definition}

\newtheorem{remark}{Remark}
\newtheorem{corollary}{Corollary}

\def\cn{\mathcal{CN}}

\usepackage{algorithm}
\usepackage{algpseudocode}
\algnewcommand\algorithmicinput{\textbf{Input:}}
\algnewcommand\INPUT{\item[\algorithmicinput]}
\algnewcommand\algorithmicoutput{\textbf{Output:}}
\algnewcommand\OUTPUT{\item[\algorithmicoutput]}

\begin{document}

\title{Coded Compressed Sensing with List Recoverable Codes for the Unsourced Random Access}

\author{Kirill Andreev\IEEEauthorrefmark{1}\thanks{\IEEEauthorrefmark{1}Kirill Andreev, Pavel Rybin and Alexey Frolov are with the Center for Next Generation Wireless and IoT (NGW), Skolkovo Institute of Science and Technology, Moscow, Russia (emails: k.andreev@skoltech.ru, p.rybin@skoltech.ru, al.frolov@skoltech.ru).}
\hspace{0.7cm} \and Pavel Rybin \IEEEauthorrefmark{1}
\hspace{0.7cm}\and Alexey Frolov\IEEEauthorrefmark{1}

\thanks{The research was carried at Skolkovo Institute of Science and Technology and supported by the Russian Science Foundation (project no. 18-19-00673), \protect\url{https://rscf.ru/en/project/18-19-00673/}}
\thanks{This paper was presented in part at 2021 Information Theory Workshop~\cite{Frolov2021List} and 2021 International Symposium on Wireless Communication Systems~\cite{Andreev2021RS}.}
}

\IEEEaftertitletext{\vspace{-2\baselineskip}}
\maketitle
\thispagestyle{plain}
\pagestyle{plain}

\begin{abstract}
We consider a coded compressed sensing approach for the unsourced random access and replace the outer tree code proposed by Amalladinne et al. with the list recoverable code capable of correcting t errors. A finite-length random coding bound for such codes is derived. The numerical experiments in the single antenna quasi-static Rayleigh fading MAC show that transition to list recoverable codes correcting t errors improves the performance of coded compressed sensing scheme by 7--10 dB compared to the tree code-based scheme. We propose two practical constructions of outer codes. The first is a modification of the tree code. It utilizes the same code structure, and a key difference is a decoder capable of correcting up to t errors. The second is based on the Reed--Solomon codes and Guruswami--Sudan list decoding algorithm. The first scheme provides an energy efficiency very close to the random coding bound when the decoding complexity is unbounded. But for the practical parameters, the second scheme is better and improves the performance of a tree code-based scheme when the number of active users is less than 200.
\end{abstract}

\section{Introduction}

The problem of massive machine-type communications (mMTC) is of critical importance for future 5G/6G wireless networks. Indeed, the number of devices connected to the network grows exponentially. At the same time, the traffic of the devices is significantly different from the traffic generated by human users and consists of short packets that are sent sporadically. The main goal is not to increase spectral efficiency but to provide connectivity and energy efficiency. Current transmission schemes are highly inefficient in this regime. The most promising way to deal with the problem is to use the random access schemes or, equivalently, a grant-free transmission, i.e., the device transmits the packet without any prior communication to the base station. As the number of devices is extremely large and it is difficult to create different encoders for the users, the promising strategy is to employ the same encoder for all the users. The receiver is not able to identify the source of the message in this case, and, thus, such schemes are called \textit{unsourced random access} (URA) schemes. The information-theoretic statement of the URA problem is proposed in \cite{polyanskiy2017perspective}. Fundamental limits and low-complexity schemes for the Gaussian MAC are given in \cite{polyanskiy2017perspective, ZPT-isit19, ordentlich2017low, vem2017user, Marshakov2019Polar, codedCS2020, Fengler2019sparcs, pradhan2019polar}. More realistic channel models such as single antenna quasi-static Rayleigh fading MAC \cite{FadingISIT2019, kowshik2019quasi, Amalladinne2019, Frolov2020ISIT} and MIMO MAC \cite{fengler2019massive} were also considered in the literature. 

This paper is inspired by a coded compressed sensing (CS) scheme proposed in \cite{amalladinne2018coupled, codedCS2020}. We note that a similar approach was already used in compressed sensing and group testing literature \cite{Cormode2006, Hung2012, Gilbert2007, Indyk2008}, but the paper \cite{codedCS2020} gives the first application of this approach for the URA problem. Clearly (see e.g.,~\cite{polyanskiy2017perspective}), the URA problem is a CS problem of huge dimensionality. The scheme from \cite{codedCS2020} utilizes the divide-and-conquer strategy, i.e., splits the task into subtasks of smaller dimensionality, solves the CS problem for each subtask, and then assembles the results. For the latter task, an outer tree code is used. We note that a similar code construction, namely a convolutional code, was used in \cite{ZigJ} but for a different single-user channel model (jamming channel or J-channel). The main drawback of the tree code is an inability to deal with errors, i.e., the codeword is not recovered if at least one of its fragments is lost. It is an actual problem for realistic channel models, such as quasi-static Rayleigh fading MAC. 

In this paper, we replace the outer tree code with the code capable of correcting $t$ errors. Note that the actual task of the outer codes is a list-recovery rather than just error correction, i.e., the decoder should recover all the codewords that are at distance at most $t$ from the channel output. Our contribution is as follows:
\begin{itemize}
    \item we derive a finite-length random coding bound for list recoverable codes correcting $t$ errors. Numerical experiments in the single antenna quasi-static Rayleigh fading MAC were carried out. The results show that transition to list recoverable codes correcting $t$ errors improves the performance of coded compressed sensing scheme by $7$--$10$ dB compared to the tree code-based scheme (the case when $t=0$). At the same time, we note that due to the restriction on the number of false messages the maximal number of active users that can be served decreases with the growth of $t$, e.g., the scheme with $t=5$ works up to $K_a \approx 400$, while the scheme with $t=0$ works up to $K_a \approx 530$ (see \Fig{fig:ebno_ka_cs}).
    \item we propose a practical scheme which is a modification of the tree code. Namely, it utilizes the same code structure, and the key difference is a decoder capable of correcting up to $t$ errors. In what follows these codes will be referred to as $t$-tree codes. We show the $t$-tree code-based scheme to provide an energy efficiency very close to the random coding bound when the decoding complexity (number of decoding paths) is unbounded. The required number of paths grows with $t$, e.g., for $K_a = 200$ the scheme with $t=0$ requires $2^8$ paths, $t=1$ requires $2^{10}$ paths, while $t=2$ requires $2^{16}$ paths (see Table~\ref{tab:LIN_paths}). 
    \item we propose Reed--Solomon (RS) code-based scheme. RS codes in combination with Guruswami--Sudan decoding algorithm are known to solve the list-recovery problem (see \cite[Section III.C]{GS} and \cite{KV}). A simple calculation of the resulting code rate shows that a straightforward application of RS codes is not possible even for the moderate number of active users. Indeed, to support $K_a$ users, we need to choose the RS code rate of less than $1/K_a$. Thus, we modify the scheme to reduce the average collision order. We start from the RS code over the smaller field and construct the codebook from several cosets of this code. For the practical parameters, this scheme is better and improves the performance of a tree code-based scheme ($t=0$) when the number of active users is less than $200$.
    \item in this paper, we focus on the single-antenna quasi-static Rayleigh fading MAC. We show that increasing $t$ is reasonable for this channel. But we also present the results for the Gaussian MAC (GMAC) in the Appendix. We show that there is no need to consider $t>0$ for this channel as the number of errors in the recovered lists is negligible. At the same time, we slightly (by $\approx 1.5$ dB) improve the CS achievability results for GMAC.   
\end{itemize}

\section{System model}
In this section, we present a system model. We need to introduce the following notations. For any positive integer $n$, we use the notation $[n] \triangleq \{1,\dots,n\}$. Let $I = \{i_1, \ldots, i_s\} \subseteq [n]$ with $i_1 < \ldots < i_s$. Given the word $\mathbf{a} = (a_1, \ldots, a_n)$, the restriction $\mathbf{a}_I$ of $\mathbf{a}$ to $I$ is the word $\mathbf{a}_I = (a_{i_1},\ldots,a_{i_s})$.

\subsection{Unsourced random access model}
Let us recall the model proposed in \cite{polyanskiy2017perspective}. We assume partial activity scenario: there are $K_\text{tot} \gg 1$ users in the system but only $K_a \ll K_\text{tot}$ are active at each time instance. Communication proceeds in a frame-synchronized fashion. The length of each frame is $n$ complex channel uses. Each active user has $k$ bits to transmit within a frame. All the users employ the same message set $[M]$ and the same codebook $\mathcal{C} = \{f(W)\}_{W=1}^M$, where $f(\cdot)$ is the encoder function. We also require $||f(W)||^2_2 \leq nP$, which means a natural power constraint.

Decoding is done up to the permutation of messages. We only require the decoder to output a set $\mathcal{L}(Y) = (W_1, W_2, \ldots, W_{K_a}) \in [M]^{K_a}$. Our main performance measures are Per User Probability of Error (PUPE)
\[
P_e = \frac{1}{K_a} \sum\limits_{i=1}^{K_a} \Pr[W_i \not\in \mathcal{L}(Y)]
\]
and False Alarm Rate (FAR)
\[
P_f = \Pr\left[ \mathcal{L}(Y) \backslash \{W_1, W_2, \ldots, W_{K_a}\} \ne \emptyset \right].
\]

\subsection{Coded compressed sensing scheme}

\begin{figure}[!ht]
    \centering
    \tikzstyle{SLOT} = [draw, rounded corners, minimum height=0.6cm, minimum width=1.45cm]
\tikzstyle{IN} =   [draw, rounded corners, minimum height=0.6cm, minimum width=1.20cm]

\tikzstyle{arrow} = [thick,->,>=stealth, dashed]
\begin{tikzpicture}[align=center, xscale = 2, yscale = 1.0]
\node [rotate=90] at (-1.5, 0.25) {Active users ($K_a$)};
\node [IN] (in1) at (-0.8, +1.5) {$W_1$};
\node [IN] (in2) at (-0.8, +0.5) {$W_2$};
\node            at (-0.8, -0.12) {$\vdots$};
\node [IN] (in3) at (-0.8, -1.0) {$W_{K_a}$};

\node (s1)  [SLOT] at (1, +1.5) {$X^{\left(1\right)}_{1}$};
\node (s2)  [SLOT] at (1, +0.5) {$X^{\left(2\right)}_{1}$};
\node (s3)  [SLOT] at (1, -1.0) {$X^{\left({K_a}\right)}_{1}$};
\node (eq1) [SLOT, rotate=0] at (1, -2.75) {$\mathbf{y}_1$};

\node [SLOT] at (2, +1.5) {$X^{\left(1\right)}_{2}$};
\node [SLOT] at (2, +0.5) {$X^{\left(2\right)}_{2}$};
\node (ss2) [SLOT] at (2, -1.0) {$X^{\left({K_a}\right)}_{2}$};
\node (eq2) [SLOT, rotate=0] at (2, -2.75) {$\mathbf{y}_2$};

\node [] at (2.75, +1.5) {$\cdots$};
\node [] at (2.75, +0.5) {$\cdots$};
\node [] at (2.75, -1.0) {$\cdots$};
\node    at (2.75, -0.12) {$\ddots$};
\node [] at (2.75, -2.75) {$\cdots$};

\node [SLOT] at (3.5, +1.5) {$X^{\left(1\right)}_{L}$};
\node [SLOT] at (3.5, +0.5) {$X^{\left(2\right)}_{L}$};
\node (ss3)[SLOT] at (3.5, -1.0) {$X^{\left({K_a}\right)}_{L}$};
\node (eq3) [SLOT, rotate=0] at (3.5, -2.75) {$\mathbf{y}_L$};

\draw [arrow] (in1) -- (s1);
\draw [arrow] (in2) -- (s2);
\draw [arrow] (in3) -- (s3);
\draw [arrow] (s3)  -- (eq1);
\draw [arrow] (ss2) -- (eq2);
\draw [arrow] (ss3) -- (eq3);

\node [fill=white,rounded corners=2pt,inner sep=1pt] at (2.25, 2.5) {Slots ($L$)};
\node [fill=white,rounded corners=2pt,inner sep=1pt] at (2.25, -1.8) {Inner code};
\node [rotate=90, fill=white,rounded corners=2pt,inner sep=1pt] at (0.1, 0.25) {Outer code};
\end{tikzpicture}
    \caption{Coded compressed sensing scheme.}
    \label{fig:ccs_scheme}
\end{figure}
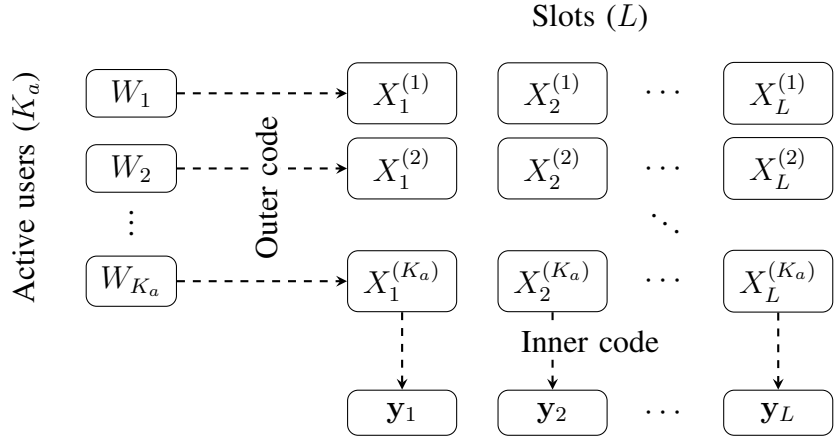

Let us briefly describe a coded compressed sensing (CS) scheme from~\cite{codedCS2020}. The transmission scheme is shown in~\Fig{fig:ccs_scheme}. The idea is to apply a divide-and-conquer strategy implemented using concatenated coding. Let us consider the $i$-th user aiming to transmit a message $W_i$. First, an outer encoder $f_O(\cdot)$ is applied, and we obtain a codeword $\ind{\bX}{i}{} = (\ind{X}{i}{1}, \ind{X}{i}{2}, \ldots, \ind{X}{i}{L}) = f_O(W_i)$, $\mathbf{X}^{(i)} \in [Q]^L$, where $L$ is both the outer code length and the number of slots (see what follows). Then the symbols $\ind{X}{i}{j}$, $j=1,\ldots,L$ are encoded with the use of inner encoder $f_I(\cdot)$. Inner code is a code over a complex field and has a length $n_1 = n/L$. The resulting codeword of the inner code is transmitted in the corresponding slot.

In this paper, we consider the single-antenna quasi-static Rayleigh fading channel and thus
\[
\mathbf{y}_j = \sum_{i=1}^{K_a} H_i f_I\left(\ind{X}{i}{j}\right) + \mathbf{z}_j
\]
where $H_i \sim \cn(0,1)$, $i=1, \ldots, K_a$ are the fading coefficients which are independent of codewords, and $\mathbf{z}_j \sim \cn(0,I_{n_1})$ is an additive white Gaussian noise (AWGN).

To recover the transmitted codewords, we first solve a CS problem for each slot. Note that the dimensionality of these problems are much smaller compared to the dimensionality of the original problem. Thus, one can use standard CS algorithms (e.g., Orthogonal Matching Pursuit, OMP, \cite{Cai2011}). See the details in Section~\ref{sec:nr}.

\begin{figure}
    \centering
    \input{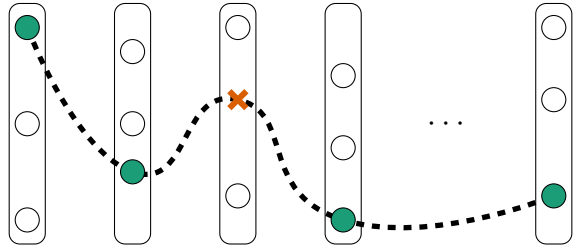}

\tikzset{
    cross/.pic = {
    \draw[rotate = 45] (-#1,0) -- (#1,0);
    \draw[rotate = 45] (0,-#1) -- (0, #1);
    }
}

\def\dX{1.4}
\def\NodeDiameter{0.16cm}
\def\yScale{1.0}
\def\dYC{\NodeDiameter * \yScale}
\def\dY{10 * \dYC}
\tikzstyle{SLOT} = [draw, rounded corners, minimum height=20*\dYC, minimum width=3*\NodeDiameter]
\begin{tikzpicture}[align=center]
\draw [line width=2pt, dashed] plot [smooth, tension=1] coordinates {(0 * \dX, 18 *\dYC) (1 * \dX, 6 *\dYC) (2 * \dX, 12 *\dYC) (3 * \dX, 2 *\dYC) (5 * \dX, 4 * \dYC)};

\node [SLOT] at (0 * \dX, \dY) {};
\node [SLOT] at (1 * \dX, \dY) {};
\node [SLOT] at (2 * \dX, \dY) {};
\node [SLOT] at (3 * \dX, \dY) {};
\node []     at (4 * \dX, \dY) {$\cdots$};
\node [SLOT] at (5 * \dX, \dY) {};

\draw[fill=\colorB] (0, 18 *\dYC) circle (\NodeDiameter);
\draw[] (0, 10 *\dYC) circle (\NodeDiameter);
\draw[] (0, 02 *\dYC) circle (\NodeDiameter);

\draw[] (\dX, 16 *\dYC) circle (\NodeDiameter);
\draw[] (\dX, 10 *\dYC) circle (\NodeDiameter);
\draw[fill=\colorB] (\dX, 06 *\dYC) circle (\NodeDiameter);

\path (2 * \dX, 12 *\dYC) pic[\colorC, line width=0.5*\NodeDiameter] {cross=1*\NodeDiameter};
\draw[] (2 * \dX, 4 *\dYC) circle (\NodeDiameter);
\draw[] (2 * \dX, 18 *\dYC) circle (\NodeDiameter);

\draw[fill=\colorB] (3 * \dX, 2 *\dYC) circle (\NodeDiameter);
\draw[] (3 * \dX,  8 *\dYC) circle (\NodeDiameter);
\draw[] (3 * \dX, 14 *\dYC) circle (\NodeDiameter);

\draw[fill=\colorB] (5 * \dX, 4 *\dYC) circle (\NodeDiameter);
\draw[] (5 * \dX, 12 *\dYC) circle (\NodeDiameter);
\draw[] (5 * \dX, 18 *\dYC) circle (\NodeDiameter);

\end{tikzpicture}
    \caption{Outer code and a list-recovery problem. A codeword which is covered by the recovered lists in all but $t=1$ positions is shown.}
    \label{fig:t_error_code}
\end{figure}

In the paper, we focus on outer code construction and decoding. After the first step, we have lists of messages (symbols of the original codewords) for each slot (see~\Fig{fig:t_error_code}). The lists may contain errors (missed and falsely detected symbols). The task of the outer code is to assemble the original codewords from the received lists. This problem is called a list-recovery problem~\cite{Guru}.

As energy efficiency is of critical importance for the mMTC scenario, our goal is to minimize the energy-per-bit ($E_b/N_0 = Pn/k$) spent by each user.

\section{Channel for the outer code}

Let us start with the case when there are no errors in the output lists. Clearly, the resulting channel is the channel without intensity information (A-channel) from \cite{Wolf1981}, which is also called a hyperchannel in the literature (see~\cite{Hyper}). Let symbols $X^{(1)}, X^{(2)}, \ldots, X^{(K_a)} \in [Q]$ be transmitted, then the output of the channel is 
\[
Y^{(A)} = \bigcup\limits_{i=1}^{K_a} X^{(i)}.
\]
The capacity of the A-channel is derived in~\cite{BasPin00}. If we consider indicator vectors of the sets then this channel can be presented as a vector OR-channel.

Our channel is a concatenation of the A-channel with the channel defined as follows. Let $Y^{(A)} \subseteq [Q]$ be the input set of messages and $Y \subseteq [Q] $ be the output set. For each element $X \in [Q]$ (the channel works on the elements independently) 
\begin{flalign*}
\pM &= \Pr[X \not\in Y | X \in Y^{(A)}],\\
\pF &= \Pr[X \in Y | X \not\in Y^{(A)}].
\end{flalign*}

Now, let us estimate the channel capacity. For simplicity, we will not perform optimization over all independent distributions of $X^{(i)}$, $i=1, \ldots, K_a$, and consider uniform distribution only. We have
\[
C_u = I(X^{(1)}, \ldots, X^{(K_a)}; Y) = H(Y) - H(Y | X^{(1)}, \ldots, X^{(K_a)}),
\]
where $X^{(i)} \overset{i.i.d.}{\sim} \mathop{unif}([Q])$ and $H(X)$ is an entropy of a random variable $X$.

Let $\Omega = |Y^{(A)}|$, clearly,
\begin{flalign*}
&H(Y | X^{(1)}, \ldots, X^{(K_a)}) = \mathbb{E}_{\Omega} [H(Y | \Omega)] \\
&= Q \left( 1 - \left(\frac{Q-1}{Q} \right)^{K_a} \right) h(\pM) + Q \left( \frac{Q-1}{Q} \right)^{K_a} h(\pF),
\end{flalign*}
where $h(x) = - x \log_2 x -(1-x)\log_2(1-x)$ is a binary entropy function.

The exact calculation of $H(Y)$ is more complicated, so we use the following estimate
\[
H(Y) \leq \sum\limits_{y \in [Q]} H\left(1_{\{y \in Y\}}\right) = Q h(\mu_{K_a}),
\]
where $1_{E}$ is an indicator of the event $E$ and
\begin{equation}\label{eq:py}
\mu_r = \left( 1-\left(\frac{Q-1}{Q}\right)^{r} \right) (1-\pM) + \left(\frac{Q-1}{Q}\right)^{r}  \pF.
\end{equation}

\begin{figure}
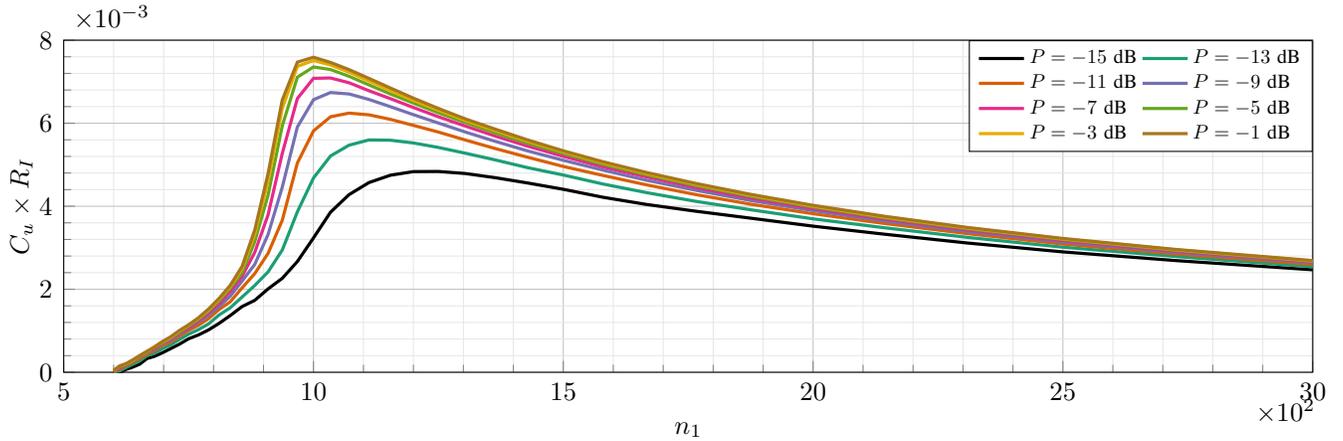

    \centering
    \input{tikz/color_scheme}
\input{tikz/capacity/double_column}

\begin{tikzpicture}
\begin{axis}[
    width=\FigureWidth, height=\FigureHeight,
    xlabel={$n_1$},
    ylabel={$C_u \times R_I$},
    xmin=500,xmax=3000,
    ymin=0, ymax=0.008,
    legend cell align={left},
    legend columns = 2,
    legend style={at={(1, 1)}, anchor=north east, font=\small, nodes={scale=0.75, transform shape}},
    axis line style={latex-latex},
    label style={font=\small},
    tick label style={font=\small},
    ylabel style={yshift=-2mm},
    grid=both,
    grid style={line width=.1pt, draw=gray!20},
    major grid style={line width=.2pt,draw=gray!50},
    tick align=inside,
    tickpos=left,
    xtick distance=500,
    minor tick num=4,
    scaled x ticks=base 10:-2,
    tick scale binop=\times,
    every x tick scale label/.style={
    at={(1,-0.08)},xshift=1pt,anchor=north east,inner sep=0pt},
]

\addplot[color=\colorA, mark=none, line width=\LineWidth] table[x=n1, y=SNR_15]{tikz/capacity/data/capacity.txt};
\addplot[color=\colorB, mark=none, line width=\LineWidth] table[x=n1, y=SNR_13]{tikz/capacity/data/capacity.txt};
\addplot[color=\colorC, mark=none, line width=\LineWidth] table[x=n1, y=SNR_11]{tikz/capacity/data/capacity.txt};
\addplot[color=\colorD, mark=none, line width=\LineWidth] table[x=n1, y=SNR_9]{tikz/capacity/data/capacity.txt};
\addplot[color=\colorE, mark=none, line width=\LineWidth] table[x=n1, y=SNR_7]{tikz/capacity/data/capacity.txt};
\addplot[color=\colorF, mark=none, line width=\LineWidth] table[x=n1, y=SNR_5]{tikz/capacity/data/capacity.txt};
\addplot[color=\colorG, mark=none, line width=\LineWidth] table[x=n1, y=SNR_3]{tikz/capacity/data/capacity.txt};
\addplot[color=\colorH, mark=none, line width=\LineWidth] table[x=n1, y=SNR_1]{tikz/capacity/data/capacity.txt};

\legend {
$P = -15$ dB,
$P = -13$ dB,
$P = -11$ dB,
$P = -9$ dB,
$P = -7$ dB,
$P = -5$ dB,
$P = -3$ dB,
$P = -1$ dB,
}
\end{axis}
\end{tikzpicture}
    \caption{Concatenated scheme rate (in bits per channel use) for $K_a = 300$ as a function of the slot length $n_1$, $R_I = c/n_1$, $c = \log_2Q = 15$ bits}
    \label{fig:capacity}
\end{figure}

We note that the estimates above are quite simple and were already presented in the literature (see~\cite{fengler2020sparcs}). We added the complete derivation for the reader's convenience. The main purpose of capacity calculation was to show (see~\Fig{fig:capacity}) that the optimal performance does not correspond to the case when we have the strongest inner code and weakest outer code (tree code). To plot these dependencies, we calculated $\pF$ and $\pM$ from inner code simulations, see the details in Section~\ref{sec:nr}. Thus, the main conclusion is that we need to use an outer list recoverable code capable of correcting $t$ errors.  

\section{Random coding bound}\label{rcb}

In this section, we derive a random coding bound for the outer code. In what follows, we consider the \textit{ensemble} of codes.

\begin{define}
Let $\mathcal{E}_1(M, L)$ be the ensemble of codebooks of size $M \times L$, where each element is sampled  i.i.d. from $\mathop{unif}([Q])$.
\end{define}

Now let us describe the decoding algorithm. Let
\[
\mathcal{Y} = (Y_1, \ldots, Y_L), \quad Y_i \subseteq [Q].
\]
Let us define list cover of $\mathcal{Y}$
\[
LC(\mathcal{Y}) = Y_1 \times \ldots \times Y_L.
\]
We require the decoder to output all the messages $W$, such that
\begin{equation}\label{dec_cond}
d(LC(\mathcal{Y}), \bX) \leq t,
\end{equation}
where $\bX = f_O(W)$, $d(LC(\mathcal{Y}), \bX) = | \{ i: X_i \not\in Y_i \}|$.



\begin{theorem}\label{thm:rcb}
There exists a code $\mathcal{C} \in \mathcal{E}_1(M, L)$, such that
\[
P_e = \sum\limits_{i=t+1}^{L} \binom{L}{i} \pM^i (1-\pM)^{L-i}
\]
and
\[
P_f \leq \sum\limits_{r=1}^{K_a} \left[\nu_r (M-r) \sum\limits_{i=0}^{t} \binom{L}{i} (1-\mu_r)^i \mu_r^{L-i} \right],
\]
where $\mu_r$ is given by eq.~\textup{(\ref{eq:py})} and
\begin{equation*}\label{eq:nur}
\nu_r = \binom{M}{r} \sum\limits_{i=0}^{r} (-1)^i \binom{r}{i} \left( \frac{r-i}{M}\right)^{K_a}.
\end{equation*}
\end{theorem}

\begin{IEEEproof}
Let us start with the false alarm rate. Let us introduce the events
\[
A_r = \{ |\{W_1, \ldots, W_{K_a}\}| = r \}, \:\: r=1,\ldots,K_a.
\]

Clearly, $\Pr[A_r] = \nu_r$. To check this one need to apply the inclusion–exclusion formula.

We have
\[
P_f = \sum\limits_{r=1}^{K_a} \nu_r \Pr\left[ \mathcal{L}(Y) \backslash \{W_1, \ldots, W_{K_a}\} \ne \emptyset | A_r \right].
\]

Let us proceed with $\Pr\left[ \mathcal{L}(Y) \backslash \{W_1, \ldots, W_{K_a}\} \ne \emptyset | A_r \right]$. W.l.o.g. assume messages $[r]$ were transmitted and let us calculate the probability that some another message $\hat{W} \in [M] \setminus [r]$ satisfy condition~(\ref{dec_cond}). Let $\hat{\bX} = f_O(\hat{W})$. Clearly, 
\[
\Pr[ \hat{X}_i \in Y_i | A_r] = \mu_r,
\]
thus the probability to accept the message $\hat{W}$ is equal to $\Pr\left[ \sum\limits_{i=1}^L \xi_i \leq t \right]$, where $\xi_i \overset{i.i.d.}{\sim} \text{Bern}(1-\mu_r)$. Applying the union bound we obtain $P_f$ from the theorem statement.

At last, note that as we utilize a single user receiver, then $P_e$ is just the probability that more than $t$ errors in the transmitted codeword have occurred.
\end{IEEEproof}

In what follows, we are interested in codebooks of size $M \approx 2^{100}$ and utilize the following upper bound for $P_f$.

\begin{corollary}
\[
P_f \leq (M-K_a) \sum\limits_{i=0}^{t} \binom{L}{i} (1-\mu_{K_a})^i \mu_{K_a}^{L-i} + p',
\]
where
\[
p' = \Pr\left[ |\{W_1, \ldots, W_{K_a}\}| < K_a \right] = 1 - \prod\limits_{i=0}^{K_a-1} \left( 1 - \frac{i}{M} \right) \leq \frac{\binom{K_a}{2}}{M}.
\]
\end{corollary}

\section{$t$-tree code-based practical scheme}

In this section, we consider and analyze a practical code construction, which is a modification of the tree code. We utilize the same code structure as in~\cite{codedCS2020}, a crucial difference is a decoder capable of correcting up to $t$ errors.

\subsection{Code construction}

Let us represent the user message $W$ as a binary $k$-bit vector $\mathbf{u}$ and split it into chunks $\mathbf{u} = (\mathbf{u}_1, \ldots, \mathbf{u}_L)$ such that $\mathbf{u}_i$ is of length $b_i$ bits, $i=1,\ldots,L$, and $\sum_{i=1}^{L} b_i = k$.

Recall that $\beg{\bu}{i} = (\bu_1, \ldots, \bu_i)$, let $B_i = \sum\nolimits_{j=1}^j b_j$. To construct the outer code, we choose the following encoding function $f_O$.
\begin{equation}\label{tree:def_nl}
X_i = f_{O, i}(\beg{\bu}{i}), \:\: i = 1, \ldots, L,
\end{equation}
where $f_{O, i}: \{0,1\}^{B_i} \to [Q]$. The main idea of the proposed code construction is that the symbol $X_i$ for the $i$-th slot depends only on the message chunks $\mathbf{u}_1, \ldots, \mathbf{u}_i$. This property allows to simplify the decoding process (see Section~\ref{tree:dec}).

Linear codes are preferred for a practical scheme, thus we construct the functions $f_{O,i}(\cdot)$, $i=1,\ldots,L$ as follows. Let $c \in \mathbb{N}$, $Q=2^c$. Let us fix a bijective mapping $\phi: \{0,1\}^{c} \to [Q]$. The major part of our construction is a binary linear code with block upper-triangular generator matrix
\begin{equation}\label{eq:G}
\mathbf{G}= 
\begin{pmatrix}
\mathbf{G}_{1,1} & \mathbf{G}_{1,2} & \mathbf{G}_{1,3} & \dots & \mathbf{G}_{1,L}\\
\mathbf{0} & \mathbf{G}_{2,2} & \mathbf{G}_{2,3} & \dots & \mathbf{G}_{2,L} \\
\mathbf{0} & \mathbf{0} & \mathbf{G}_{3,3} & \dots & \mathbf{G}_{3,L}\\
\vdots & \vdots & \vdots & \dots & \vdots \\
\mathbf{0} & \mathbf{0} & \mathbf{0} & \dots & \mathbf{G}_{L,L}\\
\end{pmatrix},
\end{equation}
where $\mathbf{G}_{j,i}$, $j = 1, \dots, L$, $i = 1, \dots, L$, is a binary matrix of size $b_{i} \times c$. 

The codeword $\bX = (X_1, \ldots, X_L) \in [Q]^L$ is obtained as follows. We start with the binary vector $\mathbf{x} = \mathbf{u} \mathbf{G}$ and then obtain a codeword $\bX$ by splitting $\mathbf{x}$ into chunks of length $c$ and applying the mapping $\phi$, i.e.
\begin{equation}\label{tree:def_l}
X_i = \phi\left( \sum\limits_{j=1}^{i} \bu_j \mathbf{G}_{j,i} \right), \:\: i = 1, \ldots, L,  
\end{equation}

\subsection{Decoding}\label{tree:dec}

Recall that the main goal of the decoder is to recover all the messages $\bu$ such that
\[
d(LC(\mathcal{Y}), \bX ) \leq t,
\]
where $\bX = f_O(\bu)$.

We note that $\bu_{[i]}$ uniquely defines $\bX_{[i]}$ for each $i=1,\ldots,L$. In what follows, we write $\bX_{[i]} = f_O(\bu_{[i]})$. This fact allows us to utilize a low-complexity decoding algorithm which decodes the blocks $\bu_i$ sequentially.

Let us introduce a notation
\[
V_l = \left\{ \bv_l \in \{0,1\}^{B_l} : d(\mathcal{Y}_{[l]}, f_O(\bv_l)) \leq t \right\}, \:\: l = 1, \ldots, L,
\]
which means the list of messages at each of the decoding steps.

\begin{algorithm}
\caption{Decoding algorithm}\label{alg:dec}
\begin{algorithmic}[1]
\INPUT $\mathcal{Y}$
\OUTPUT $V_L$ \Comment Decoded messages
\State $V_0 \gets \emptyset$
\For {$l = 1, \ldots, L$} 
\State $V_l \gets \emptyset$ 
\For {$\bv_{l-1} \in V_{l-1}$} \Comment For each element from the list
\For {$\bu_l \in \{0,1\}^{b_l}$} \Comment For each next block
\State $\bv_l \gets (\bv_{l-1}, \bu_l)$
\If { $d(\mathcal{Y}_{[l]}, f_O(\bv_l)) \leq t$ }
\State $V_l \gets V_l \bigcup \bv_l$
\EndIf 
\EndFor
\EndFor
\EndFor
\State \Return $V_L$
\end{algorithmic}
\end{algorithm}

\begin{remark}
As we see, Algorithm~\ref{alg:dec} is guaranteed to recover the transmitted message in case no more than $t$ errors have occurred. In what follows, we analyze false alarm rate and the complexity. Note that the complexity depends on $|V_l|$, $l=1,\ldots, L$.  
\end{remark}

\subsection{Analysis}


Let $M_l = 2^{B_l}$, we start with calculating $\mathbb{E}[|V_l|]$ for $1 \leq l \leq L$.  We apply random coding for the following ensemble.

\begin{define}
The elements of the ensemble $\mathcal{E}_2(b_1, \ldots, b_L, L)$ are obtained by random choice of the generator matrix $\mathbf{G}$ with the structure defined by eq.~\textup(\ref{eq:G}\textup), i.e., each non-zero element is sampled i.i.d. from $\text{Bern}(1/2)$ distribution.
\end{define}

\begin{lemma}
The following bounds hold for the ensemble $\mathcal{E}_2(b_1, \ldots, b_L, L)$
\begin{equation}\label{eq:vl}
\mathbb{E}[|V_l|] \leq \overline{v}_l \triangleq M_l \sum\limits_{j=0}^l \rho_j \lambda_j, \:\: l = 1, \ldots, L,
\end{equation}
where
\begin{flalign}\label{eq:Pl}
&\lambda_j = \left(1 - \frac{1}{M_{j + 1}}\right)^{K_a} - \left(1 - \frac{1}{M_{j}}\right)^{K_a}, \:\: j=1,\ldots,l-1, \nonumber\\
&\lambda_0 = \left(1 - \frac{1}{M_{1}}\right)^{K_a}, \nonumber\\
&\lambda_l = 1 - \left(1 - \frac{1}{M_{l}}\right)^{K_a}
\end{flalign}
and
\[
\rho_j = \sum\limits_{\begin{array}{c} 0 \leq x \leq j\\ 0 \leq y \leq l-j\\ x+y \leq t\end{array}} \binom{j}{x} \binom{l-j}{y} \pM^x (1-\pM)^{j-x} 
\gamma_1^{y} \gamma_2^{l-j-y}, 
\]
where
\[
\gamma_1 = \left( \frac{K_a}{Q} \right) \pM + \left(1 - \frac{1}{Q} \right)(1-\pF)
\]
and
\[
\gamma_2 =  \left( \frac{K_a}{Q} \right) (1-\pM) + \left(1 - \frac{1}{Q} \right)\pF
\]
\end{lemma}

\begin{IEEEproof}
Assume that messages $\ind{\bu}{1}{}, \ind{\bu}{2}{}, \ldots, \ind{\bu}{K_a}{}$ were transmitted. Consider some another information word $\hat{\bu}$ and calculate the probability $\Pr\left[ d(\mathcal{Y}_{[l]}, \hat{\bX}_{[l]}) \leq t \right]$, where $\hat{\bX}_{[l]} = f_O(\hat{\bu}_{[l]})$. In what follows, we assume $\hat{\bu}$ to be fixed while $\ind{\bu}{1}{}, \ind{\bu}{2}{}, \ldots, \ind{\bu}{K_a}{}$ to be chosen uniformly at random.

The main difference compared to the proof of Theorem~\ref{thm:rcb} is as follows. The beginning of the information word $\hat{\bu}$ may coincide with the beginning of one of the transmitted information words. In this case the beginnings of the codewords will also coincide, which should be taken into account in the analysis.

Let us introduce the events
\[
E_{u,j} = \{ \beg{\hat{\bu}}{j} = \ind{\bu}{u}{[j]} \}, \:\: u = 1, \ldots, K_a.
\]

Let $E^c$ be a complementary event to $E$, let
\[
E_j = \left( \bigcup\limits_{u=1}^{K_a} E_{u,j} \right) \bigcap \left(\bigcap\limits_{u=1}^{K_a} E^c_{u,j+1} \right)
\]
be the event that the longest match length is equal to $j$.

Clearly, the probability $\lambda_j = \Pr\left[ E_j \right]$ is given by eq.~(\ref{eq:Pl}).

Now consider
\[
\rho_j = \Pr\left[ d(\hat{\bX}_{[l]}, \mathcal{Y}_{[l]}) \leq t | E_j \right]. 
\]

Consider the slots $1, \ldots, j$. Clearly, $\Pr\left[\hat{X}_l \in Y^{(A)}_l | E_j \right] = 1$ for $l=1,\ldots,j$. Thus, we can only have an error in case of miss-detection, and the result can be described by i.i.d. random variables $\xi_i \sim \text{Bern}(p_m)$.

Consider the slots $j+1, \ldots, L$. Note that 
\[
\frac{1}{Q} \leq \Pr\left[\hat{X}_l \in Y^{(A)}_l | E_j \right] \leq \frac{K_a}{Q}, \:\: l=j+1, \ldots, L,
\]
and thus, 
\[
\Pr\left[\hat{X}_l \not\in Y_l | E_j \right] \leq \gamma_1,
\]
and
\[
\Pr\left[\hat{X}_l \in Y_l | E_j \right] \leq \gamma_2,
\]
for $l=j+1, \ldots, L$.

\end{IEEEproof}

\begin{theorem}\label{thm:linear_bound}
There exists a code $\mathcal{C} \in \mathcal{E}_2(b_1, \ldots, b_L, L)$, such that
\[
P_e = \sum\limits_{i=t+1}^{L} \binom{L}{i} \pM^i (1-\pM)^{L-i}
\]
and
\[
P_f \leq \overline{v}_L,
\]
where $\overline{v}_L$ is given by eq.~\textup{(\ref{eq:vl})}.
\end{theorem}

\section{Reed-Solomon code-based practical scheme}

Reed--Solomon codes in combination with Guruswami--Sudan decoding algorithm are known to solve the list-recovery problem. In this section, we develop a Reed-Solomon code-based practical scheme.

\subsection{Reed--Solomon codes}
Let $\mathbb{F}_Q$ be the field with $Q$ elements and let $\mathbb{F}_Q[X]$ denote the ring of polynomials over $\mathbb{F}_Q$. Let $\beta_1, \beta_2, \ldots, \beta_L \in \mathbb{F}_Q$ and $\beta_i \ne \beta_j$ when $i \ne j$. We define an $[n_O = L, k_O]$ Reed--Solomon code $\mathcal{C}$ as follows
\begin{flalign*}
\mathcal{C} &= \left\{ ( f(\beta_1), f(\beta_2), \ldots, f(\beta_L) ) : \right. \\
&\left. f(x) \in \mathbb{F}_Q[X],  \deg{f(x)} < k_O \right\}
\end{flalign*}

\subsection{Naive approach}
In this section, we are to apply Guruswami--Sudan list recovery algorithm to our problem in a straightforward manner. In what follows, we briefly explain the idea and refer the reader to \cite{PPM3} for the details. Let us enumerate the elements of the field $\mathbb{F}_Q$ in some order as follows
\[
\mathbb{F}_Q = \{\alpha_1, \alpha_2, \ldots, \alpha_Q\}.
\]

For now, let us present $\mathcal{Y}$ as a binary matrix (indicator matrix) $\mathbf{Y} = [y_{i,j}]$ of size $Q\times L$ as follows: $y_{i,j} = 1$ iff $\alpha_i \in Y(j)$.

Let us define a $Q \times L$ matrix $\mathbf{M} = [m_{i,j}]$ of multiplicities. We set $m_{i,j} = 0$ if $y_{i,j} = 0$. When $y_{i,j} = 1$, we can select $m_{i,j}$ to be any positive integer number. In what follows, we use the following matrix
\[
\mathbf{M} = m \mathbf{Y}
\]
for some positive integer $m$.

Given the multiplicity matrix, we can apply the Guruswami--Sudan decoding algorithm (see Algorithm~\ref{alg:GS}).

\begin{algorithm}
\caption{Decoding algorithm}\label{alg:GS}
\begin{algorithmic}[1]
\INPUT $L$, $k_O$, set of locators, matrix of multiplicities $\mathbf{M}$
\OUTPUT List of polynomials $f(x)$

\State \textit{Interpolation}. Find a bivariate polynomial $Q(x, y)$ of minimal $(1,k-1)$-weighted degree that passes thought each point $(\beta_j, \alpha_i)$, $i=1,\ldots,Q$, $j=1,\ldots,L$, with multiplicity $m_{i,j}$.  
\State \textit{Factorization}. Find all the factors of $Q(x; y)$ of type $y - f(x)$ with $\deg{f(x)} < k_O$.
\end{algorithmic}
\end{algorithm}

Let
\[
C(\mathbf{M}) = \frac{1}{2} \sum\limits_{i=1}^Q  \sum\limits_{j=1}^L m_{i,j} (m_{i,j}+1). 
\]

The Algorithm~\ref{alg:GS} is known (see \cite{GS, KV}) to include $f(x)$ in the output list if
\begin{equation}\label{rec_cond}
(\mathbf{M}, \mathbf{C}) \geq \sqrt{2(k_O-1)C(\mathbf{M})},
\end{equation}
where $(\cdot , \cdot)$ is a dot product of matrices and the matrix $\mathbf{C} = [c_{i,j}]$ is a matrix corresponding to the codeword $c = (f(\beta_1, \ldots, f(\beta_L))$, i.e. $c_{i,j} = 1$ iff $\alpha_i = f(\beta_j)$. Clearly, this matrix have only one unit in each column. 

The following upper bound holds for the list size
\[
L(\mathbf{M}) \leq \sqrt{\frac{2 C(\mathbf{M})}{k_O-1}}.
\]

Let us investigate the recovery condition~(\ref{rec_cond}) in more details. Assume that in each position we have lists of size $K_a$, then we have
\[
m (L - d(\mathcal{Y}, c)) \geq  \sqrt{(k_O-1) m(m+1)K_a L}
\]
and thus we can recover the codeword if the number of errors ($t =  d(\mathcal{Y}, c)$) satisfies the inequality
\[
t \leq L \left( 1 - \sqrt{\frac{(k_O-1)}{L} K_a \frac{(m+1)}{m}} \right).
\]

As we see, to have error-correcting capabilities, we have to require the outer code rate $R_O \triangleq k_O/L \leq 1/K_a$, which is infeasible even for a moderate number of users. Thus, our next goal is to reduce the average number of collisions.

\subsection{Modified scheme}
\label{sec:rs_modified}

We propose to consider the $[L,k_O]$ Reed--Solomon code $\mathcal{C}_0$ over a smaller field $\mathbb{F}_q$ and construct a common codebook by using several cosets of this code, i.e.
\[
\mathcal{C} = \bigcup_{u=0}^{S-1} \{ \mathbf{v}_u + \mathcal{C}_0\}, \quad \mathbf{v}_u \in \mathbb{F}_Q^L.
\]
In what follows, we use $\mathbf{v}_u = p_u (q,q, \ldots, q)$, $p_u = 0, \ldots, Q/q-1$.

Consider a frame of length $n$ consisting of $L$ slots. Each slot consists of $\log_2Q$ information bits which are encoded by inner code having length $n_1 = n / L$.
\begin{figure}[!htbp]
    \centering
    \def\dX{1.8cm}
\def\dY{1.5cm}
\input{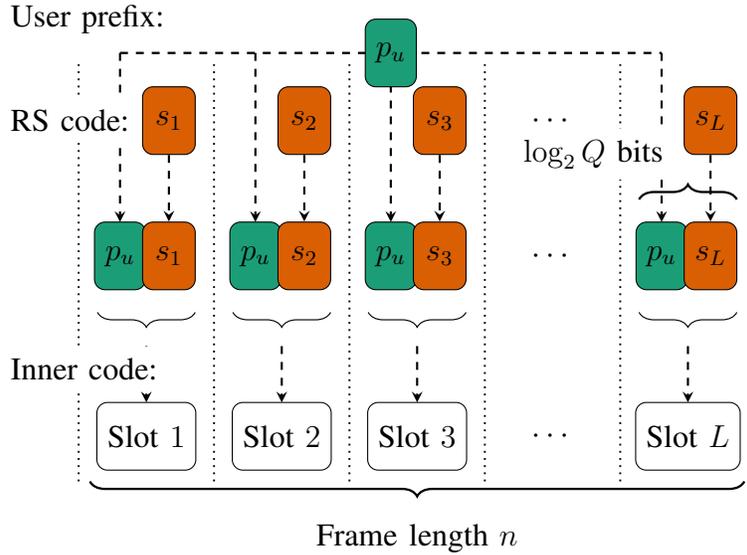}
\tikzstyle{PREFIX} = [draw, fill=\colorB, rounded corners, minimum height=0.6 * \dY, minimum width=0.6cm]
\tikzstyle{QRS} = [draw, fill=\colorC, rounded corners, minimum height=0.6 * \dY, minimum width=0.7cm]
\tikzstyle{IN} =   [draw, rounded corners, minimum height=0.6 * \dY, minimum width=1.20cm]
\tikzstyle{SLOT} = [draw, rounded corners, minimum height=0.6 * \dY, minimum width=1.3cm]
\tikzstyle{arrow} = [thick,->,>=stealth, dashed]
\begin{tikzpicture}[align=center]

\node (s1) [QRS] at (0 * \dX + 0.65cm, 1.2 * \dY) {$s_1$};
\node (s2) [QRS] at (1 * \dX + 0.65cm, 1.2 * \dY) {$s_2$};
\node (s3) [QRS] at (2 * \dX + 0.65cm, 1.2 * \dY) {$s_3$};
\node []         at (3 * \dX + 0.35cm, 1.2 * \dY) {$\cdots$};
\node (s5) [QRS] at (4 * \dX + 0.65cm, 1.2 * \dY) {$s_L$};

\node () [PREFIX] at (0 * \dX,          0) {$p_u$};
\node (ss1) [QRS] at (0 * \dX + 0.65cm, 0) {$s_1$};

\node () [PREFIX] at (1 * \dX,          0) {$p_u$};
\node (ss2) [QRS] at (1 * \dX + 0.65cm, 0) {$s_2$};

\node () [PREFIX] at (2 * \dX,          0) {$p_u$};
\node (ss3) [QRS] at (2 * \dX + 0.65cm, 0) {$s_3$};

\node []          at (3 * \dX + 0.35cm, 0) {$\cdots$};

\node () [PREFIX] at (4 * \dX,          0) {$p_u$};
\node (ss5) [QRS] at (4 * \dX + 0.65cm, 0) {$s_L$};

\draw [arrow] (s1) -- (ss1);
\draw [arrow] (s2) -- (ss2);
\draw [arrow] (s3) -- (ss3);
\draw [arrow] (s5) -- (ss5);

\draw[decorate,decoration={brace,amplitude=2mm}, line width=0.3mm] (-0.3cm + 4 * \dX, 0.5 * \dY) -- (1.0cm + 4 * \dX, 0.5 * \dY);

\draw[decorate,decoration={brace,amplitude=2mm, mirror}, line width=0.15mm] (-0.3cm + 0 * \dX,-0.5 * \dY) -- (1.0cm + 0 * \dX, -0.5 * \dY);
\draw[decorate,decoration={brace,amplitude=2mm, mirror}, line width=0.15mm] (-0.3cm + 1 * \dX,-0.5 * \dY) -- (1.0cm + 1 * \dX, -0.5 * \dY);
\draw[decorate,decoration={brace,amplitude=2mm, mirror}, line width=0.15mm] (-0.3cm + 2 * \dX,-0.5 * \dY) -- (1.0cm + 2 * \dX, -0.5 * \dY);
\draw[decorate,decoration={brace,amplitude=2mm, mirror}, line width=0.15mm] (-0.3cm + 4 * \dX,-0.5 * \dY) -- (1.0cm + 4 * \dX, -0.5 * \dY);

\draw [dotted, line width=0.25mm] (-1 * \dX / 2 + 0.35cm, -2.0 * \dY) -- (-1 * \dX / 2 + 0.35cm, 1.7 * \dY);
\draw [dotted, line width=0.25mm] ( 1 * \dX / 2 + 0.35cm, -2.0 * \dY) -- ( 1 * \dX / 2 + 0.35cm, 1.7 * \dY);
\draw [dotted, line width=0.25mm] ( 3 * \dX / 2 + 0.35cm, -2.0 * \dY) -- ( 3 * \dX / 2 + 0.35cm, 1.7 * \dY);
\draw [dotted, line width=0.25mm] ( 5 * \dX / 2 + 0.35cm, -2.0 * \dY) -- ( 5 * \dX / 2 + 0.35cm, 1.7 * \dY);
\draw [dotted, line width=0.25mm] ( 7 * \dX / 2 + 0.35cm, -2.0 * \dY) -- ( 7 * \dX / 2 + 0.35cm, 1.7 * \dY);
\draw [dotted, line width=0.25mm] ( 9 * \dX / 2 + 0.35cm, -2.0 * \dY) -- ( 9 * \dX / 2 + 0.35cm, 1.7 * \dY);

\draw [arrow] (0 * \dX + 0.35cm, -0.8* \dY) -- (0 * \dX + 0.35cm, -1.3* \dY);
\draw [arrow] (1 * \dX + 0.35cm, -0.8* \dY) -- (1 * \dX + 0.35cm, -1.3* \dY);
\draw [arrow] (2 * \dX + 0.35cm, -0.8* \dY) -- (2 * \dX + 0.35cm, -1.3* \dY);
\draw [arrow] (4 * \dX + 0.35cm, -0.8* \dY) -- (4 * \dX + 0.35cm, -1.3* \dY);

\node [SLOT] at (0 * \dX + 0.35cm, -1.6* \dY) {Slot $1$};
\node [SLOT] at (1 * \dX + 0.35cm, -1.6* \dY) {Slot $2$};
\node [SLOT] at (2 * \dX + 0.35cm, -1.6* \dY) {Slot $3$};
\node []     at (3 * \dX + 0.35cm, -1.6* \dY) {$\cdots$};
\node [SLOT] at (4 * \dX + 0.35cm, -1.6* \dY) {Slot $L$};

\draw[line width=0.3mm, dashed] (0, 1.8 * \dY) -- (4 * \dX, 1.8 * \dY);
\draw[arrow] (0 * \dX, 1.8 * \dY) -- (0 * \dX, 0.3* \dY);
\draw[arrow] (1 * \dX, 1.8 * \dY) -- (1 * \dX, 0.3* \dY);
\draw[arrow] (2 * \dX, 1.8 * \dY) -- (2 * \dX, 0.3* \dY);
\draw[arrow] (4 * \dX, 1.8 * \dY) -- (4 * \dX, 0.3* \dY);
\node () [PREFIX] at (2 * \dX, 1.8 * \dY) {$p_u$};

\node [fill=white, anchor=west] at (-1.6cm, -1.0 * \dY) {Inner code:};
\node [fill=white, anchor=west] at (-1.6cm,  1.2 * \dY) {RS code:};
\node [fill=white, anchor=west] at (-1.6cm,  2.1 * \dY) {User prefix:};
\node [fill=white]  at (3.3 * \dX + 0.35cm,  0.9 * \dY) {$\log_2{Q}$ bits};

\draw[decorate,decoration={brace,amplitude=2mm, mirror}, line width=0.3mm] (-0.4cm,-2.0 * \dY) -- (1.1cm + 4 * \dX ,-2.0 * \dY);
\node [fill=white]  at (2 * \dX + 0.35cm, -2.5 * \dY) {Frame length $n$};

\end{tikzpicture}
    \caption{Frame structure with slots consisting of Reed-Solomon code symbols and spreading prefix. Slot decoding procedure follows the inner-code decoding rules.}
    \label{fig:rs_scheme}
\end{figure}
Each user transmits $k$ information bits followed by $h$ CRC bits (required to suppress the false alarm rate below the required threshold). We use the Reed--Solomon (RS) code to encode these bits. As we discovered earlier, an RS code requires a small collision number. To reduce the collision number in each slot, we need to spread $K_a$ users using a prefix $p_u$ (See~\Fig{fig:rs_scheme}). The first $x_p$ information bits of each slot correspond to the same prefix value $p_u \sim U\left(0, 2^{x_p} - 1\right)$ generated by the user for each transmission. The user copies the same value of the prefix into each slot. The remaining $\log_2q = \log_2{Q} - x_p$ bits are devoted to the RS code symbols. As the slot count is $L$, the length of the RS code (constructed over the field $\mathbb{F}_q$) is also $L$. Then, the resulting $\log_2{Q}$ bits are encoded by the inner code and transmitted over the slot length $n_1$ channel uses with the total frame length $n$.

\section{Numerical results}\label{sec:nr}
\begin{figure}[t]
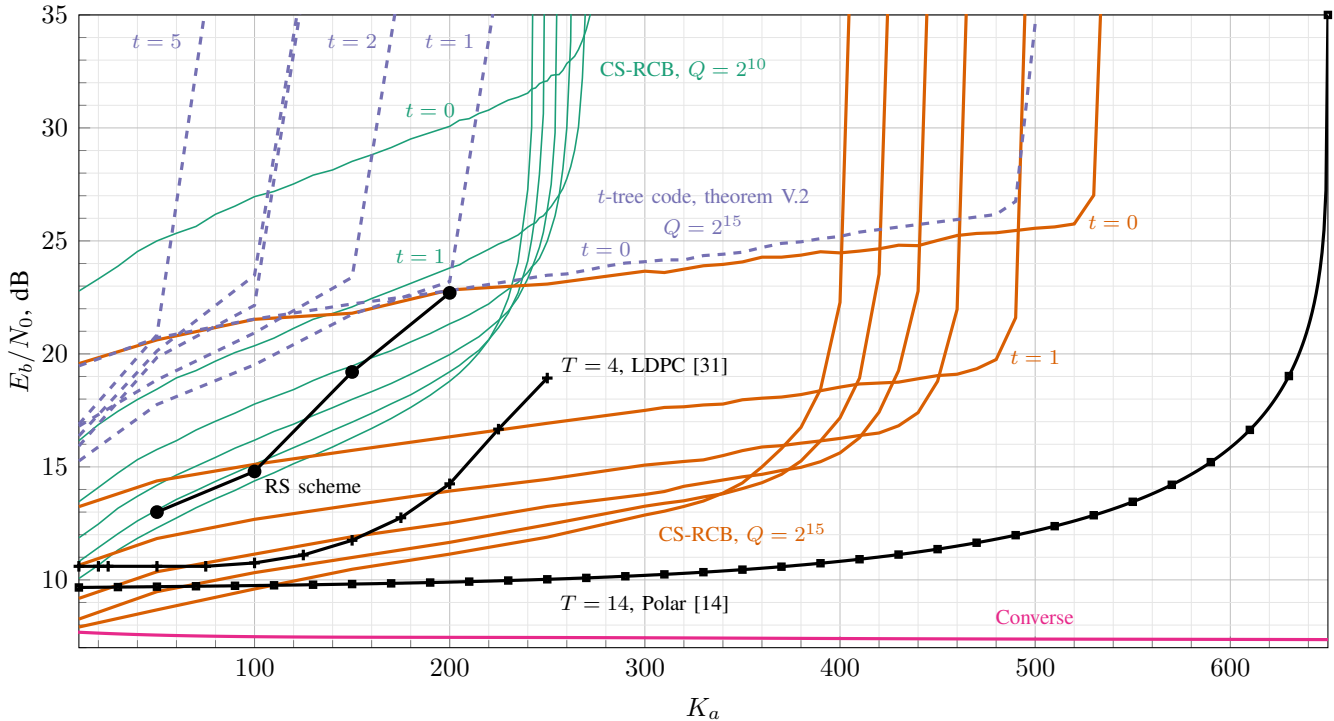

\centering
\input{tikz/color_scheme}
\input{tikz/ebno_ka/double_column}

\begin{tikzpicture}[trim axis left, trim axis right]
\begin{axis}[
    width=\FigureWidth, height=\FigureHeight,
    xlabel={$K_a$},
    ylabel={${E_b}/{N_0}$, dB},
    xmin=10,xmax=650,
    ymin=7, ymax=35.0,
    legend cell align={left},
    legend columns = 1,
    legend style={at={(0, 1)}, anchor=north west, font=\small},
    axis line style={latex-latex},
    label style={font=\small},
    tick label style={font=\small},
    ylabel style={yshift=-2mm},
    grid=both,
    grid style={line width=.1pt, draw=gray!20},
    major grid style={line width=.2pt,draw=gray!50},
    tick align=inside,
    tickpos=left,
    xtick distance=100,
    minor tick num=4,
    x tick label style={
        /pgf/number format/.cd,
        fixed,
        fixed zerofill,
        precision=0
    }
]

\newcommand{\PlotRCB}[1]{
\addplot table[x=Ka, y=t0]{#1};
\addplot[forget plot] table[x=Ka, y=t1]{#1};
\addplot[forget plot] table[x=Ka, y=t2]{#1};
\addplot[forget plot] table[x=Ka, y=t3]{#1};
\addplot[forget plot] table[x=Ka, y=t4]{#1};
\addplot[forget plot] table[x=Ka, y=t5]{#1};
}

\begin{scope}[color=\colorB, mark=none, solid, line width=0.5 * \LineWidth]
\PlotRCB{tikz/ebno_ka/data/ebno_ka_ks10_N1.txt}
\node[above] at (320, 31.6) {\scriptsize{CS-RCB, $Q=2^{10}$}};
\node[above] at (190, 30.0) {\scriptsize{$t=0$}};
\node[above] at (185, 23.5) {\scriptsize{$t=1$}};
\end{scope}

\begin{scope}[color=\colorC, mark=none, solid, line width=\LineWidth]
\PlotRCB{tikz/ebno_ka/data/ebno_ka_ks15_N1.txt}
\node[above] at (350, 11.0) {\scriptsize{CS-RCB, $Q=2^{15}$}};
\node[above] at (540, 25.0) {\scriptsize{$t = 0$}};
\node[above] at (500, 19.0) {\scriptsize{$t = 1$}};
\end{scope}

\begin{scope}[color=\colorD, mark=none, dashed, line width=\LineWidth]
\PlotRCB{tikz/ebno_ka/data/ebno_ka_ks15_N1_conv.txt}
\node[above] at (330, 26.0) {\scriptsize{$t$-tree code, theorem~\ref{thm:linear_bound}}};
\node[above] at (330, 24.6) {\scriptsize{$Q=2^{15}$}};
\node[above] at (50, 33) {\scriptsize{$t=5$}};

\node[above] at (150, 33) {\scriptsize{$t=2$}};
\node[above] at (200, 33) {\scriptsize{$t=1$}};
\node[above] at (280, 23.8) {\scriptsize{$t=0$}};
\end{scope}

\begin{scope}[color=\colorA, mark=*, mark size = 2, line width = \LineWidth]
\addplot table[x=Ka, y=EBNO]{tikz/ebno_ka/data/RS.txt};
\node[above] at (130, 13.3) {\scriptsize{RS scheme}};
\end{scope}

\begin{scope}[color=\colorA, mark=square, mark size = 1.0, mark repeat={20}, solid, line width=\LineWidth]
\addplot table[x=Ka, y=EBNO]{tikz/ebno_ka/data/polar_T14.txt};
\node[above] at (300, 8.0) {\scriptsize{$T=14$, Polar~\cite{Frolov2020ISIT}}};
\end{scope}
\begin{scope}[color=\colorA, mark=none, mark=+, solid, line width=\LineWidth]
\addplot table[x=Ka, y=EBNO]{tikz/ebno_ka/data/aloha4_ldpc.txt};
\node[above] at (300, 18.5) {\scriptsize{$T=4$, LDPC~\cite{Kowshik2020TCOM}}};
\end{scope}
\begin{scope}[color=\colorE, mark=none, solid, line width=\LineWidth]
\addplot table[x=Ka, y=EBNO]{tikz/ebno_ka/data/converse.txt};
\node[above] at (500, 7.5) {\scriptsize{Converse}};
\end{scope}

\end{axis}
\end{tikzpicture}
\caption{Numerical results for the coded compressed sensing scheme for $Q=2^{10}$ and $Q=2^{15}$, $t=0, \ldots, 5$. The parameters are as follows: Rayleigh fading channel, $k=100$ bits, $n = 30000$, $P_e=10^{-1}$, $P_f=10^{-3}$, $K_0$ and $n_1$ are chosen to minimize the required $E_b/N_0$. $T$-fold ALOHA ($T=4$ and $T=14$) is added as a reference. Reed-Solomon practical solution has parameters taken from the Table~\ref{tab:RA_params}.}
\label{fig:ebno_ka_cs}
\end{figure}

\begin{table}
\centering
\begin{tabular}{|r|cccccc|}
\hline
$K_a$ & $t = 0$ & $t = 1$ & $t = 2$ & $t = 3$ & $t = 4$ & $t = 5$\\
\hline
$50$ &  $12 \ (0.556) $ & $14 \ (0.476) $ & $15 \ (0.444) $ & $16 \ (0.417) $ & $18 \ (0.370) $ & $19 \ (0.351) $ \\
$100$ & $14 \ (0.476) $ & $15 \ (0.444) $ & $17 \ (0.392) $ & $18 \ (0.370) $ & $19 \ (0.351) $ & $20 \ (0.333) $ \\
$150$ & $15 \ (0.444) $ & $16 \ (0.417) $ & $18 \ (0.370) $ & $19 \ (0.351) $ & $20 \ (0.333) $ & $21 \ (0.317) $ \\
$200$ & $16 \ (0.417) $ & $17 \ (0.392) $ & $19 \ (0.351) $ & $20 \ (0.333) $ & $21 \ (0.317) $ & $22 \ (0.303) $ \\
$250$ & $16 \ (0.417) $ & $18 \ (0.370) $ & $19 \ (0.351) $ & $21 \ (0.317) $ & $22 \ (0.303) $ & $23 \ (0.290) $ \\
\hline
\end{tabular}
\caption{Optimal slot count (and outer code rate $R_O$) for random coding bound in the Rayleigh fading channel, $Q=2^{15}$, and $n=30000$\label{tab:RCB_slots}.}
\end{table}

Let us consider a communication system with the frame length $n = 30000$ channel uses. Each user transmits $k=100$ bits within the frame performing outer encoding with $Q$-ary code and the inner encoding with a random spherical codebook having $Q$ codewords. Each symbol of the outer code is being transmitted within a slot having length $n_1 = n/L$.

For the inner code, we use a randomly generated spherical codebook (of length $n_1$). We use a codebook with codewords having an i.i.d. uniform distribution on the (complex) power shell. We decode the inner code using the OMP~\cite{Cai2011} and its MMSE-based extension~\cite{Sparrer2016}. As the OMP is a sequential algorithm, the number of output codewords equals the number of steps. Thus, we pass an additional parameter $K_0$ -- the output list size for a given slot. By varying the $K_0$ value, one can change the balance between $\pF$ and $\pM$.

Let us start our analysis with the compressed sensing-based random coding bound (CS-RCB) from Theorem~\ref{thm:rcb}. Results for the random coding bound in the single-antenna Rayleigh fading channel are presented in~\Fig{fig:ebno_ka_cs} by green thin lines for $Q=2^{10}$, and by orange lines for $Q=2^{15}$ for $t=0, \ldots, 5$. To find an optimal performance, we need to find such a pair of $K_0$ and $n_1$ that deliver the minimum $E_b/N_0$ for some number of active users $K_a$ such that per-user probability of error $P_e < 0.1$ and the false alarm rate $P_f < 10^{-3}$.

We evaluated the CS-RCB as follows. The $P_e$ depends on $K_0$ and $n_1$ (or, equivalently, on $L$). To describe the optimization procedure, let us consider some signal-to-noise ratio value. One must find the minimum $P_e$ over all possible $K_0$ and $L$ values. To solve this problem, let us first perform the optimization over $K_0$. To do this, let us fix the slot count and evaluate the receiver operating characteristic (parametrized by $K_0$) and find the minimum $P_e$ such that $P_f / P_e < 10^{-2}$ for some slot count $L$ and the signal-to-noise ratio. The sequential nature of the OMP decoder significantly simplifies the ROC-curve construction: one needs to set the decoder list size to be sufficiently large and then evaluate $\pM$ and $\pF$ for the whole $K_0$ range. Next, we found that the $P_e$ has a single minimum over the slot count $L$ at any fixed signal-to-noise ratio. Indeed, the slot count decrease weakens the outer code, while the slot count increase weakens the inner code performance. As a result, to find the optimal slot count, one must check that the $P_e$ at neighbor slot values is higher. The final step is to perform this optimization for different $E_b/N_0$ and find the minimum value at which $P_e < 0.1$ and $P_f < 10^{-3}$ by testing different signal-to-noise ratios.

We evaluated the CS-RCB for $t=0, \ldots, 5$ (see~\Fig{fig:ebno_ka_cs}). As a reference, we added the converse bound from~\cite{Kowshik2020TCOM}, the $4$-fold LDPC-based ALOHA from~\cite{FadingISIT2019} and $14$-fold ALOHA with polar codes from~\cite{Frolov2020ISIT} to~\Fig{fig:ebno_ka_cs}. One can observe a significant (more than $10$ dB for $t=5$ and $K_a = 50$) improvement compared to $t=0$ case. When the number of active users is small, the scheme with $t=5$ demonstrates better energy efficiency than a $T$-fold ALOHA with polar codes from~\cite{Frolov2020ISIT}, known as the best practical solution for the fading channel with a single antenna at the receiver. When the number of active users grows, the CS-RCB for higher values of $t$ ``saturates''\footnote{Saturation means that the minimum $E_b/N_0$ goes to $\infty$} faster ($t=5$ saturates at $K_a\approx 400$, while $t=0$ works up to $K_a \approx 530$). The growth of the parameter $t$ requires more slots and lower outer code rate. The optimal slot count values $L$ and the outer code rate $R_O = \frac{k}{L\log_2Q}$ for different $t$ of the CS-RCB are presented in the Table~\ref{tab:RCB_slots}.

\begin{table}
\centering
\begin{tabular}{|r|cccccc|}
\hline
$K_a$ & $t = 0$ & $t = 1$ & $t = 2$ & $t = 3$ & $t = 4$ & $t = 5$\\
\hline
$50$ &  $13 \ (0.513) $ & $21 \ (0.317) $ & $33 \ (0.202) $ & $46 \ (0.145) $ & $54 \ (0.123) $ & $64 \ (0.104) $ \\
$100$ & $14 \ (0.476) $ & $25 \ (0.267) $ & $40 \ (0.167) $ & $54 \ (0.123) $ & $66 \ (0.101) $ & -- \\
$150$ & $15 \ (0.444) $ & $30 \ (0.222) $ & $47 \ (0.142) $ & -- & -- & -- \\
$200$ & $16 \ (0.417) $ & $34 \ (0.196) $ & -- & -- & -- & -- \\
$250$ & $17 \ (0.392) $ & -- & -- & -- & -- & -- \\
\hline
\end{tabular}
\caption{Optimal slot count (and outer code rate $R_O$) achievability bound for the code from Theorem~\ref{thm:linear_bound} in the Rayleigh fading channel, $Q=2^{15}$, and $n=30000$, and maximum average paths $\mathbb{E}[|V_l|] \leq 2^{10}$\label{tab:LIN_slots}.}
\end{table}
To evaluate the achievability bound for the tree code from Theorem~\ref{thm:linear_bound} ($t$-tree code), one can use the same procedure as the CS-RCB minimum $E_b/N_0$ search. The first step is to find the minimum $P_e$ that satisfies the false alarm rate constraints and then find the minimum $E_b/N_0$ over different slot counts. The main difference compared to the CS-RCB is as follows. When finding the minimum $P_e$, one must satisfy the constraint on the maximum number of decoding paths ($v^\star$). We choose $v^\star = 2^{10}$ to plot the curves.

Thus the optimization problem can be formulated as follows.
$$
\text{Minimize} \ E_b/N_0, \quad \text{subject} \ \mathbb{E}[|V_l|] \leq v^\star,  \ l = 1, \ldots, L, \ \sum_{l=1}^L b_l = k, \ P_e < 0.1, \ P_f < 10^{-3}.
$$
To do this, we utilized a greedy information bits allocation for each slot every time we evaluated the $P_e$ and $P_f$. This procedure starts from the first slot and assigns the maximum number of information bits to each slot keeping the average number of decoding paths below some threshold. If the total number of assigned bits becomes smaller than $k=100$, we assume $P_e=1$. The resulting energy efficiency is presented in~\Fig{fig:ebno_ka_cs} by blue dashed lines for $t=0, \ldots, 5$. The outer coding rates are presented in~Table~\ref{tab:LIN_slots}. We note that the energy efficiency for the Theorem~\ref{thm:linear_bound} bound and for the CS-RCB are almost the same for $t=0$ case. For $t=5$ this difference becomes dramatic: the energy efficiency becomes much worse and the ``saturation'' happens at $K_a \approx 100$ for the Theorem~\ref{thm:linear_bound} bound. The main cause of this behavior is the $\mathbb{E}[|V_l|] \leq v^\star$ constraint. To limit the number of paths, our greedy information bits allocation algorithm is unable to assign many bits to each subsequent slot. This limit requires more slots, making the inner code weaker. Resulting split of $k=100$ information bits among slots is presented in~Table~\ref{tab:LIN_bits} for $t=0$ and $t=1$. We have validated these optimal configurations via simulations and confirmed the resulting energy efficiency and the false alarm rate.

\begin{table}
\centering
\begin{tabular}{|r|l|}
\hline
\multicolumn{1}{|c}{$K_a$} & \multicolumn{1}{|c|}{Information bits pattern}\\
\hline
\multicolumn{2}{|c|}{$t=0$} \\
\hline
$50$ & $ \mathbf{b} = \left[15\ 11\ 8\ 9\ 8\ 9\ 8\ 9\ 8\ 9\ 6\ 0\ 0\right]$ \\
$100$ & $\mathbf{b} = \left[15\ 9\ 8\ 8\ 8\ 8\ 8\ 8\ 8\ 8\ 8\ 4\ 0\ 0\right]$ \\
$150$ & $\mathbf{b} = \left[15\ 8\ 8\ 7\ 7\ 7\ 8\ 7\ 8\ 7\ 8\ 7\ 3\ 0\ 0\right]$ \\
$200$ & $\mathbf{b} = \left[15\ 7\ 7\ 7\ 6\ 7\ 7\ 7\ 7\ 7\ 7\ 7\ 7\ 2\ 0\ 0\right]$ \\
\hline
\multicolumn{2}{|c|}{$t=1$} \\
\hline
$50$ & $\mathbf{b} = \left[9\ 4\ 4\ 4\ 4\ 4\ 4\ 4\ 4\ 4\ 4\ 4\ 4\ 8\ 7\ 8\ 8\ 8\ 4\ 0\ 0\right]$ \\
$100$ & $\mathbf{b} = \left[9\ 3\ 3\ 3\ 3\ 3\ 3\ 3\ 3\ 3\ 3\ 3\ 3\ 3\ 3\ 3\ 7\ 7\ 7\ 8\ 7\ 7\ 3\ 0\ 0\right]$ \\
$150$ & $\mathbf{b} = \left[9\ 2\ 2\ 2\ 2\ 2\ 2\ 2\ 2\ 2\ 2\ 2\ 2\ 3\ 2\ 3\ 3\ 3\ 3\ 3\ 3\ 6\ 7\ 7\ 7\ 7\ 7\ 3\ 0\ 0\right]$ \\
$200$ & $\mathbf{b} = \left[9\ 2\ 2\ 2\ 2\ 2\ 2\ 2\ 2\ 2\ 2\ 2\ 2\ 2\ 2\ 2\ 2\ 2\ 2\ 2\ 2\ 2\ 2\ 2\ 6\ 6\ 7\ 6\ 7\ 6\ 7\ 0\ 0\ 0\right]$ \\
\hline
\end{tabular}
\caption{Greedy bit allocation results for the Rayleigh fading channel, $v^\star = 2^{10}$\label{tab:LIN_bits}.}
\end{table}

\begin{table}
\centering
\begin{tabular}{|l|cccccc|}
\hline
& $t = 0$ & $t = 1$ & $t = 2$ & $t = 3$ & $t = 4$ & $t = 5$ \\
\hline
CS-RCB                & 22.7 & 16.3 & 13.9 & 12.5 & 11.7 & 11.1 \\ 
$v^\star = \infty$ & 22.7 & 16.4 & 13.9 & 12.6 & 11.7 & 11.3 \\ 
$v^\star = 2^{16}$ & 22.7 & 16.9 & 16.2 & 16.8 & 17.6 & -- \\ 
$v^\star = 2^{10}$ & 22.7 & 23.2 & -- & -- & -- & -- \\ 
$v^\star = 2^{8}$ & 24.4 & -- & -- & -- & -- & -- \\ 
\hline
\end{tabular}
\caption{Energy efficiency ($E_b/N_0$) for different restrictions on the maximal number of paths ($v^\star$), $K_a = 200$\label{tab:LIN_paths}.}
\end{table}
Moreover, if one sets the maximum number of paths $v^\star = \infty$, the resulting energy efficiency coincides with the CS-RCB for all $t=0, \ldots, 5$ and for all $K_a$ values that we evaluated, and the optimal slot count does not change compared to the CS-RCB. To evaluate the performance degradation given some decrease in $v^\star$ constraint, we have evaluated the energy efficiency for different maximum sizes $v^\star$ at $K_a=200$ (See Table~\ref{tab:LIN_paths}).

\begin{table}
\centering
\begin{tabular}{|r|ccccc|}
\hline
$K_a$ & $K_0$ & $L$ & $x_p$ & $R_{RS}$ & $h$\\
\hline
$50$ & $54$ & $40$ & $7$ & $0.3250$ & $14$\\
$100$ & $120$ & $40$ & $8$ & $0.3750$ & $14$\\
$150$ & $180$ & $45$ & $8$ & $0.3333$ & $15$\\
$200$ & $240$ & $42$ & $9$ & $0.4048$ & $15$\\
\hline
\end{tabular}
\caption{System parameters for Reed-Solomon-based solution\label{tab:RA_params}}
\end{table}
Next, we have evaluated the RS-based practical solution. For the RS scheme, we need to optimize the following parameters for each number of active users $K_a$: the inner-code list size $K_0$, the length of the user prefix $x_p$ (bits), and the number of slots in the frame $L$. We also need to adjust the number of CRC bits $h$ to suppress false detection, but, during the simulations, we just evaluated this value and corrected the $E_b/N_0$ value presented in~\Fig{fig:ebno_ka_cs} by the number of additional parity check bits. The Reed-Solomon coding rate $R_{RS}$ mentioned in the Table~\ref{tab:RA_params} also does not include the value $b$ ($R_O = R_{RS} R_{CRC}$, where $R_{CRC} = k/(k+h)$).

When searching for the optimal RS-code parameters, we performed an ad-hoc optimization of $K_0$, $x_p$, and $L$ parameters for our practical solution and did not scan the full parameter space. Nevertheless, we can make the following conclusions. The increase of $K_a$ requires both a longer prefix and a larger RS code length. These two requirements are actually in contradiction. Indeed, the RS code is constructed over the field of size $q = Q / 2^{x_p}$, and $q > L$. On the other hand, the prefix length increase results in the decrease of $q$, and one cannot increase the number of slots in the frame. The slot count increase weakens the inner-code performance. We also note, that the practical RS-based scheme operates over a larger outer code length compared to the CS-RCB.

\section{Conclusions and future work}

In this paper, we proposed to use list recoverable codes correcting $t$ errors in the coded compressed sensing scheme. We have derived the finite-length random coding bound for such codes and evaluated the energy efficiency of the resulting CS scheme in the single-antenna base station in the Rayleigh fading channel. The results show that transition to list recoverable codes correcting $t$ errors improves the performance of coded compressed sensing scheme by $7$--$10$ dB compared to the tree code-based scheme (the case when $t=0$). We propose two practical constructions of outer codes. The first one is a modification of the tree code. It utilizes the same code structure, and the major difference is a decoder capable of correcting up to $t$ errors. The second one is based on the Reed--Solomon codes and Guruswami--Sudan list decoding algorithm. The first scheme provides an energy efficiency very close to the random coding bound when the decoding complexity (i.e. the number of decoding paths) is unbounded. But for the practical parameters (the number of decoding paths is restricted with $v^* = 2^{10}$), the second scheme is better and improves the performance of a tree code-based scheme ($t=0$) when the number of active users is less than $200$. At the same time, both practical schemes are far from the random coding bound for $t=5$, and we pose the construction of good list recoverable codes with low-complexity decoding as an open question. Another interesting further research directions are a) improving the random coding bound, e.g., by considering a multi-user reception, which should reduce the false alarm rate significantly; b) the use of soft or semi-soft information provided by the inner code decoder in the outer code.

\bibliographystyle{IEEEtranTCOM}
\bibliography{main}
\appendices

\section{Coded compressed sensing for the Gaussian MAC}
\begin{figure}[ht!]
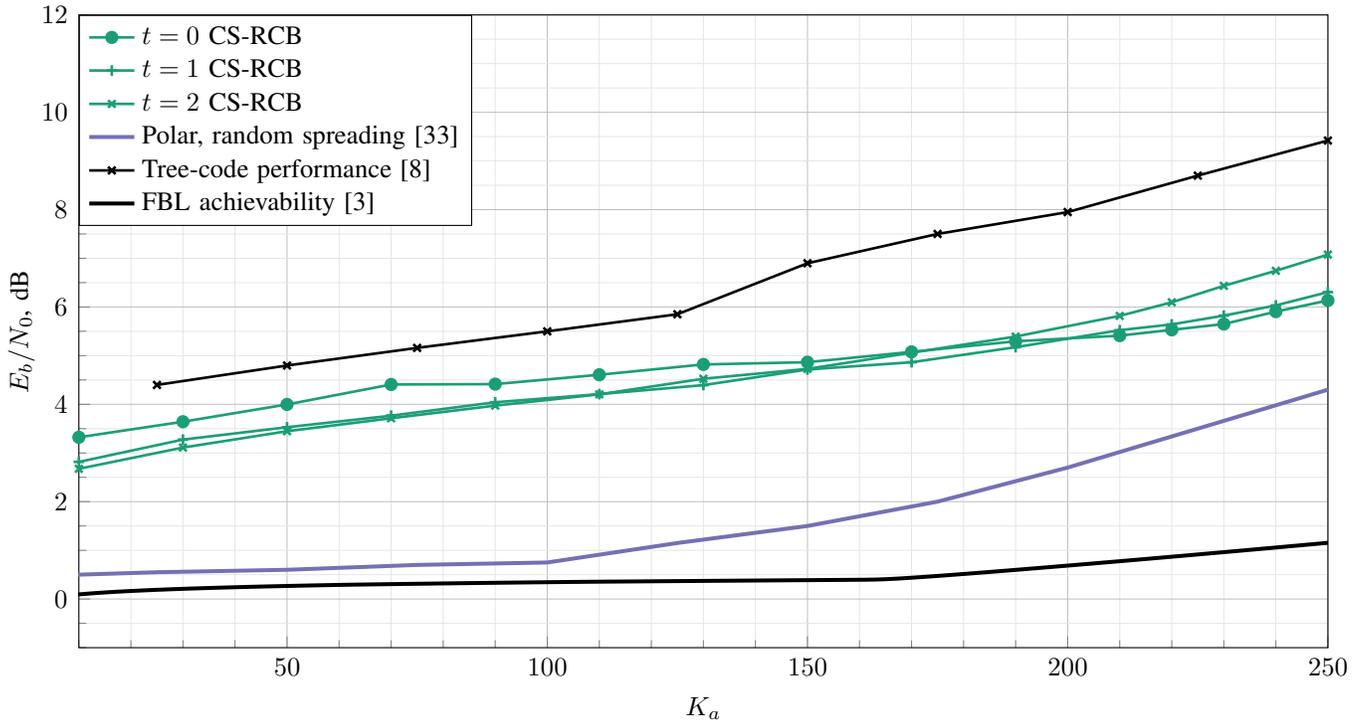

\centering
\input{tikz/color_scheme}
\input{tikz/ebno_ka/double_column}

\begin{tikzpicture}[trim axis left, trim axis right]
\begin{axis}[
    width=\FigureWidth, height=\FigureHeight,
    xlabel={$K_a$},
    ylabel={${E_b}/{N_0}$, dB},
    xmin=10,xmax=250,
    ymin=-1, ymax=12.0,
    legend cell align={left},
    legend columns = 1,
    legend style={at={(0, 1)}, anchor=north west, font=\small},
    axis line style={latex-latex},
    label style={font=\small},
    tick label style={font=\small},
    ylabel style={yshift=-2mm},
    grid=both,
    grid style={line width=.1pt, draw=gray!20},
    major grid style={line width=.2pt,draw=gray!50},
    tick align=inside,
    tickpos=left,
    xtick distance=50,
    minor x tick num=4,
    minor y tick num=3,
    x tick label style={
        /pgf/number format/.cd,
        fixed,
        fixed zerofill,
        precision=0
    }
]


\begin{scope}[color=\colorB, line width=1, mark size=2]
\addplot[mark=*] table[x=Ka, y=t0]{tikz/ebno_ka_awgn/data/ebno_ka_ks15_AWGN.txt};
\addplot[mark=+] table[x=Ka, y=t1]{tikz/ebno_ka_awgn/data/ebno_ka_ks15_AWGN.txt};
\addplot[mark=x] table[x=Ka, y=t2]{tikz/ebno_ka_awgn/data/ebno_ka_ks15_AWGN.txt};
\end{scope}

\addplot[color=\colorD, line width=1.5] table[x=Ka, y=EBNO]{tikz/ebno_ka_awgn/data/random_spread.txt};

\addplot[color=\colorA, mark=x, line width=1.0] table[x=Ka, y=EBNO]{tikz/ebno_ka_awgn/data/treecode.txt};
\addplot[color=\colorA, mark=none, line width=1.5] table[x=Ka, y=EBNO]{tikz/ebno_ka_awgn/data/ach_mac_raw.txt};

\legend {
$t=0$ CS-RCB,
$t=1$ CS-RCB,
$t=2$ CS-RCB,
{Polar, random spreading~\cite{AmalladinnePolar2020}},
Tree-code performance~\cite{codedCS2020},
FBL achievability\cite{polyanskiy2017perspective},
}

\end{axis}

\end{tikzpicture}
\caption{Performance of coded compressed sensing scheme for $Q=2^{10}$ and $Q=2^{15}$, $t=0, \ldots, 5$. The parameters are as follows: AWGN channel, $k=100$ bits, $n = 30000$, $P_e=P_f = 5\times 10^{-2}$, $K_0$ and $n_1$ are chosen to minimize the required $E_b/N_0$. Random coding bound for the outer code is used.}
\label{fig:ebno_ka_cs_awgn}
\end{figure}
In the paper, we focus on the single antenna quasi-static Rayleigh fading MAC. We show that increasing $t$ is reasonable for this channel. The reader may ask if there is an improvement in the Gaussian MAC when $t>0$. We present the results for GMAC in this section. We have evaluated the CS-RCB and present the results in \Fig{fig:ebno_ka_cs_awgn}. We have also added results from~\cite{codedCS2020}, the finite-blocklength (FBL) achievability bound from~\cite{polyanskiy2017perspective} and the results from~\cite{AmalladinnePolar2020}, which, to the best of the authors' knowledge, outperform all existing practical schemes in the range $K_a \leq 250$. We use the same OMP algorithm (without channel estimation step) as in the Rayleigh fading channel model. We need to point out that the difference between $t=0$ and $t=5$ cases for the GMAC becomes much smaller. Moreover, there is no reason to construct the outer code able to correct more than $t>3$ errors. Thus, we conclude that there is no need to consider $t>0$ for this channel as the number of errors in the recovered lists is negligible. At the same time, we slightly (by $\approx 1.5$ dB) improve the CS achievability results for GMAC.
\end{document}